\RequirePackage{ifpdf}
\ifpdf 
\documentclass[pdftex]{sigma}
\else
\documentclass{sigma}
\fi

\begin{document}
\allowdisplaybreaks
\renewcommand{\PaperNumber}{090}

\FirstPageHeading

\renewcommand{\thefootnote}{$\star$}

\ShortArticleName{Non-Commutative Mechanics in Mathematical \& in
Condensed Matter Physics}

\ArticleName{Non-Commutative Mechanics in Mathematical\\ \& in
Condensed Matter Physics\footnote{This paper is a contribution to
the Proceedings of the O'Raifeartaigh Symposium on
Non-Perturbative and Symmetry Methods in Field Theory
 (June 22--24, 2006, Budapest, Hungary).
The full collection is available at
\href{http://www.emis.de/journals/SIGMA/LOR2006.html}{http://www.emis.de/journals/SIGMA/LOR2006.html}}}

\Author{Peter A.~HORV\'ATHY}
\AuthorNameForHeading{P.A.~Horv\'athy}

\Address{Laboratoire de Math\'ematiques et de Physique
Th\'eorique, Universit\'e de Tours,\\  Parc de Grandmont,  F-37200
Tours, France}
\Email{\href{mailto:horvathy@lmpt.univ-tours.fr}{horvathy@lmpt.univ-tours.fr}} 

\ArticleDates{Received September 25, 2006, in f\/inal form
November 27, 2006; Published online December 14, 2006}

\Abstract{Non-commutative structures were introduced,
 independently and around the same time, in mathematical and in condensed matter phy\-sics
  (see Table~1).
Souriau's construction applied to the two-parameter central
extension of the planar Galilei group leads to the ``exotic''
particle, which has non-commuting position coordinates. A
Berry-phase argument applied to the Bloch electron yields in turn
a semiclassical model
 that has
been used to explain the anomalous/spin/optical Hall ef\/fects.
The non-commutative parameter is momentum-dependent in this case,
and can take the form of a monopole in momentum space.}


\Classification{81V70; 81T75}

\section[''Exotic'' Galilean symmetry and mechanics in the plane]{``Exotic''
 Galilean symmetry and mechanics in the plane}

Central extensions f\/irst entered physics when Heisenberg
realized that, in the quantum mechanics of a massive
non-relativistic particle, the position and momentum operators did
not commute. As a consequence, phase-space translations act only
up-to-phase
 on the quantum Hilbert space.  In  more mathematical terms, it
 is not the [commutative] translation group itself, only its
[non-commutative] $1$-parameter central extension, the Heisenberg
group, which is represented unitarily. Similarly, Galilean boosts
act, for a massive non-relativistic system, only up-to phase. In
other words, it is the $1$-parameter central extension of the
Galilei group, called the Bargmann group, that acts unitarily.
True representations only arise for massless particles.

Are there further extension parameters? In $d\geq3$ space
dimensions, the Galilei group admits a $1$-parameter central
extension only \cite{Barg}. The extension parameter, $m$, is
identif\/ied with the physical mass. However, L\'evy-Leblond
\cite{LL}  recognized that, in the plane, the Galilei group admits
a second  extension, highlighted by the non-commutativity of the
Galilean boost generators,
\begin{gather*}
[K_1,K_2]=i\kappa,
\end{gather*}
where $\kappa$ is the new extension parameter.
 This has long been considered, however, a mere mathematical
curiosity, as planar physics  has itself been viewed a toy. The
situation started to change around 1995, though, with the
construction of physical models which realize this ``exotic''
symmetry
 \cite{Grigore,Grigore-a,Grigore-b,Grigore-c,LSZ,LSZ-a}.
These models have the strange feature that the Poisson bracket of
the planar coordinates does not vanish,
\begin{gather*}
\{x_1,x_2\}=\frac{\kappa}{m^2}\equiv\theta.
\end{gather*}

{\small\begin{table}[thp] \centering \caption{Exotic Galilean
symmetry vs.\ semiclassical models with Berry term.}
\vspace{1mm}
\label{tableau}
\begin{tabular}{|l|l|}
\hline HIGH-ENERGY/MATH. PHYS.
&CONDENSED MATTER PHYS.\\
\hline 1970 L\'evy-Leblond: &1983 Laughlin: theory of FQHE
\\
``exotic''  planar
Galilei group&\\
\hline 1995--97 Duval, Grigore, Brihaye, Lukierski
&1995--2000 Niu et al.: Berry term \\
mechanical models with exotic symmetry &for semiclassical Bloch
electron
\\
\hline 2000--2001 Duval et al.: &2002--2003
Jungwirth--Niu--MacDonald:
\\
exotic particle in e.m. f\/ield \& Hall ef\/fect; &Anomalous Hall
ef\/fect
\\
non-commutative mechanics
&\\
\hline 2004 B\'erard, Mohrbach: &2003 Fang et al.: monopole in
\\ momentum dependent monopole-type
&momentum space in Anomalous Hall ef\/fect
\\
 non-commutativity&2003 Murakami--Nagaosa--Zhang:
\\
& spin-Hall ef\/fect
\\
\hline 2000 Jackiw--Nair exotic structure from &2005 Sinova et al:
\\
relativistic spin& observation of spin-Hall ef\/fect
\\
\hline 2005 Duval et al: ``SpinOptics'' &2004
Onoda--Murakami--Nagaosa,
\\
& Bliokh: Optical Magnus/Hall ef\/fect
\\
\hline
\end{tabular}
\end{table}}

\renewcommand{\thefootnote}{\arabic{footnote}}
\setcounter{footnote}{0}

\section{The exotic model}

In \cite{Grigore,Grigore-a,Grigore-b,Grigore-c,DH,DH-a,DH-b,DH-c}
Souriau's ``orbit method'' \cite{SSD} was used to construct a
classical planar system associated with L\'evy-Leblond's two-fold
extended
 Galilean symmetry.  It has an ``exotic''
symplectic form and a free Hamiltonian,
\begin{gather*}
\Omega_0=dp_i\wedge dx_i+
\frac{1}{2}\theta\,\epsilon_{ij}\,dp_i\wedge{}dp_j,
\qquad H_0=\frac{\vec{p}\,{}^2}{2m}.
\end{gather*}
The associated (free) motions follow the usual straight lines; the
``exotic'' structure behaves, roughly, as spin: it enters the
(conserved) boost and the angular momentum,
\begin{gather*}
j=\epsilon_{ij}x_ip_j +\frac{\theta}{2}\,{\vec{p}}\,{}^2, \qquad
K_{i}=mx_{i}-p_{i}t+m\theta\,\epsilon_{ij}p_{j}.
\end{gather*}
The new terms are separately conserved, though. The new structure
does not seem, hence, to lead to any new physics.

The situation changes dramatically, though, if the particle is
coupled to a gauge f\/ield. Souriau's minimal coupling
prescription \cite{SSD} yields indeed
\begin{gather*}
\Omega=\Omega_0+eB\,dx_1\wedge dx_2, \qquad H=H_0+eV.
\end{gather*}
The  associated Poisson bracket automatically satisf\/ies the
Jacobi identity; equations of motion read
\begin{gather}
m^*\dot{x}_{i}=p_{i}-\displaystyle
em\theta\,\epsilon_{ij}E_{j},\nonumber \qquad
\dot{p}_{i}=eE_{i}+eB\,\epsilon_{ij}\dot{x}_{j}, \label{DHeqmot}
\end{gather}
where $\theta=k/m^2$ is the non-commutative parameter and
\begin{gather}
m^*=m(1-e\theta B). \label{effmass}
\end{gather}

The novel features, crucial for physical applications, are
two-fold. They both concern the f\/irst equation in
(\ref{DHeqmot}).

Firstly, the interplay between the exotic structure and the
magnetic f\/ield yields the \textit{effective mass} $m^*$ in
(\ref{effmass}).

\begin{figure}[t]
\centerline{\includegraphics[scale=0.5]{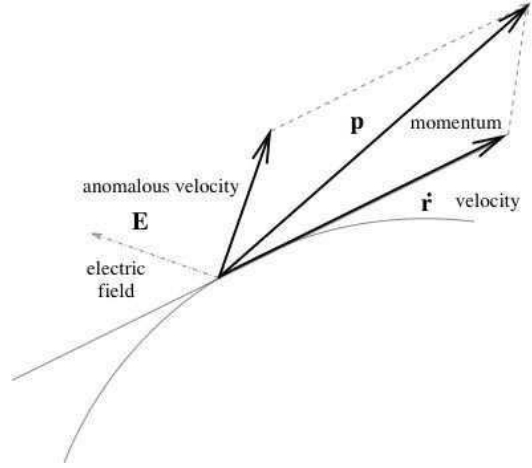}}\vspace{-3mm}

\caption{Due to the anomalous velocity term perpendicular to the
electric f\/ield, the velocity and the momentum may be
non-parallel.}\label{velmom}
\end{figure}

Secondly, the  \textit{anomalous velocity term}, perpendicular to
the direction of the electric f\/ield, makes that velocity and
momentum, $\dot{x}_i$ and $p_{i}$, are not  parallel in general, cf.~Fig.~\ref{velmom}.

Such a possibility has been discarded by some high-energy
physicists. However, it has been argued a long time ago
\cite{Dixon,Souriau74,Souriau74-a}, that {\it no first principle
requires that velocity and momentum be proportional}, and that
relaxing this restriction allows for perfectly consistent
theories. This has been well-known in condensed matter physics,
where the velocity is $\partial\epsilon/\partial{\boldsymbol p}$
where the band energy,  $\epsilon$, may  dif\/fer from the simple
quadratic expression ${\boldsymbol p}^2/2m$. The novelty is the
additional ``anomalous'' velocity term, see
Section~\ref{BlochSect} below.

Equations (\ref{DHeqmot}) derive from the f\/irst-order
``phase-space'' Lagrangian
\begin{gather}
 \int\!({\boldsymbol p}-{\boldsymbol A}\,)\cdot d{\boldsymbol x}-\frac{p^2}{2}\,dt
+\frac{\theta}{2}\,{\boldsymbol p}\times d{\boldsymbol p}.
\label{exolag}
\end{gather}

When $m^*\neq0$, (\ref{DHeqmot}) is also a Hamiltonian system,
$\dot{\xi}=\{h,\xi_\alpha\}$, with $\xi=(p_i,x_j)$ and Poisson
brackets
\begin{gather*}
\{x_{1},x_{2}\}=\frac{m}{m^*}\,\theta, \qquad
\{x_{i},p_{j}\}=\frac{m}{m^*}\,\delta_{ij}, \qquad
\{p_{1},p_{2}\}=\frac{m}{m^*}\,eB.
\end{gather*}

A remarkable property is that for \textit{vanishing effective
mass} $m^*=0$, i.e., when the magnetic f\/ield takes the critical
value $B={1}/({e\theta})$, the system becomes singular. Then
``Faddeev--Jackiw'' (alias symplectic) reduction yields an
essentially two-dimensional, simple system, reminiscent of
``Chern--Simons mechanics'' \cite{DJT,DJT-a}\footnote{The model
goes back to \cite{DJT-aa}, see \cite{DJT-b}.}. The symplectic
plane plays, simultaneously, the role of both conf\/iguration and
phase space.

The clue is to introduce the ``twisted'' coordinate
\begin{gather*}
{\boldsymbol Q}= {\boldsymbol r}-{\boldsymbol q},\qquad
{\boldsymbol r}=(x_i),\qquad q_i=\epsilon_{ij}\frac{p_j}{eB}.
\end{gather*}
In the critical case $eB\theta=1$ the momentum stops to be
dynamical: it is determined by the position according to
\begin{gather*}
p_i=m\epsilon_{ij}\frac{E_j}{B}.
\end{gather*}
Then ${\boldsymbol Q}$ becomes the  guiding center, ${\boldsymbol
Q}={\boldsymbol r}+{m{\boldsymbol E}}/({eB^2})$. The reduced
system has Poisson bracket \& energy
\begin{gather*}
\{Q_1,Q_2\}_{\rm red}=\frac{1}{eB}, \qquad H_{\rm
red}=eV(Q_1,Q_2)+\frac{\theta^2e^2m}{2}{\boldsymbol E}^2.
\end{gather*}
${\boldsymbol Q}$ follows therefore a generalized Hall law.
 The ``rotating coordinate''
${\boldsymbol q}={\boldsymbol r}-{\boldsymbol Q}$ becomes in turn
``frozen'' to the ${\boldsymbol q}=-{m{\boldsymbol E}}/{eB^2}$,
determined (via the electric f\/ield) by the position alone. The
evolution of ${\boldsymbol r}$ is hence rigidly determined by that
of ${\boldsymbol Q}$.

Quantization of the reduced system yields the wave functions
Laughlin starts with~\cite{QHE}\footnote{On the quantum Hall
ef\/fect see, e.g., \cite{QHE-a}.}.

Let us now illustrate our general theory on examples.

\begin{figure}[t]
\centerline{\includegraphics[scale=.5]{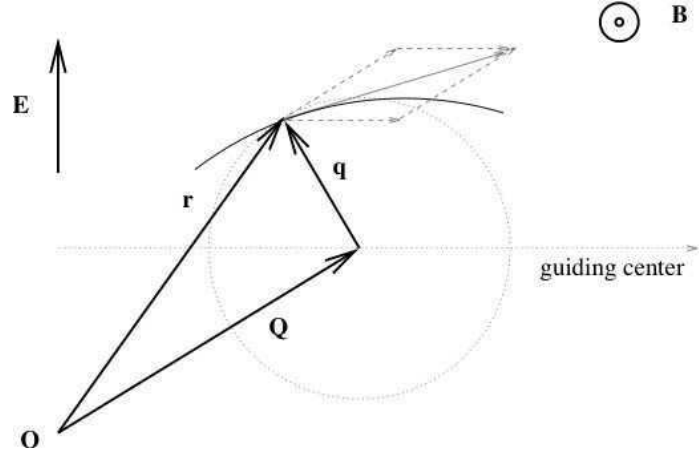}}\vspace{-3mm}

\label{CFIELD1} \caption{Composition of the uniform guiding center
motion and of uniform rotation around it yields the usual
cyclotronic motion.}
\end{figure}

\begin{figure}[t]
\centerline{\includegraphics[scale=.45]{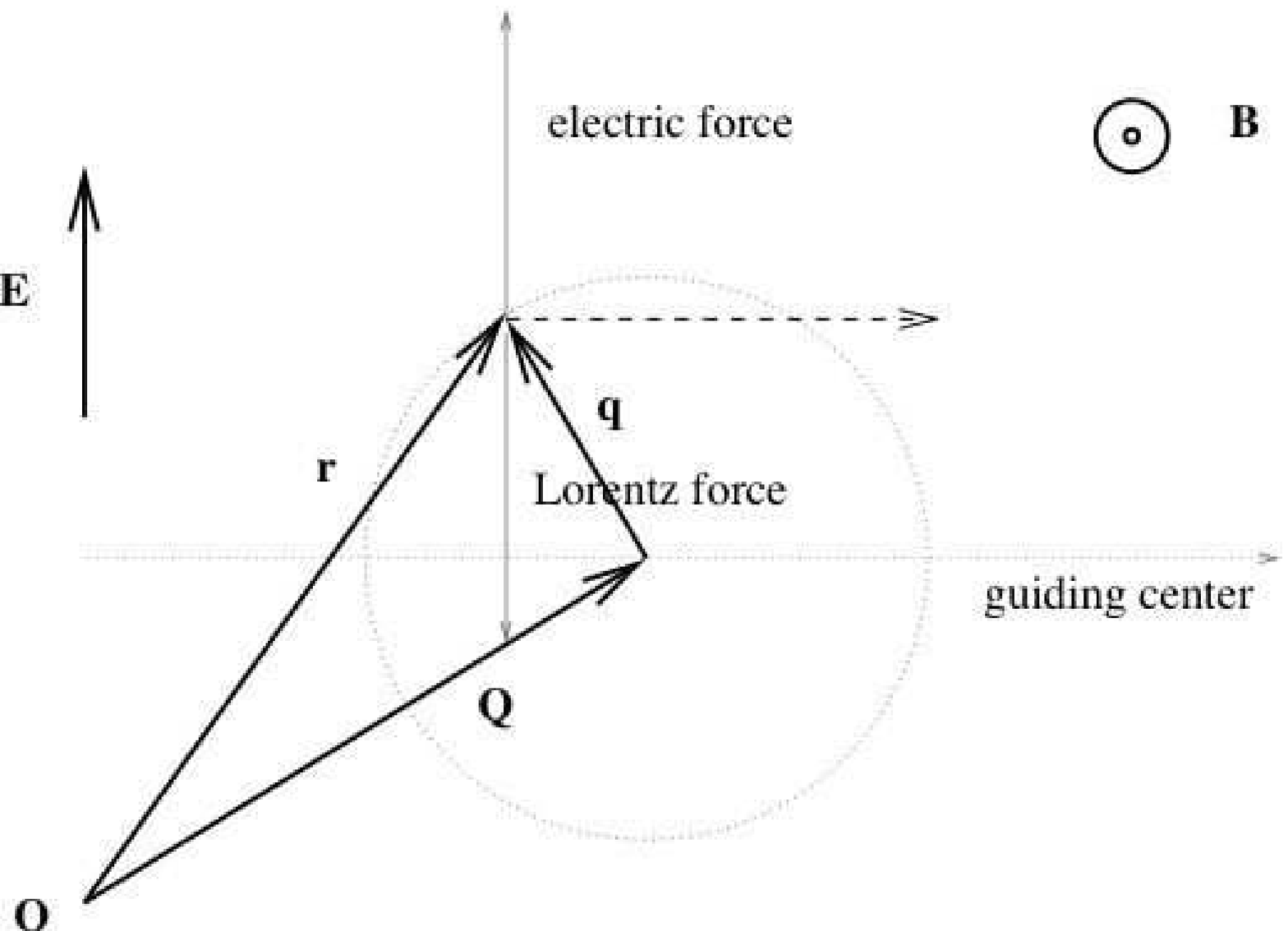}
\hspace{10pt plus 1fil}%
\includegraphics[scale=.45]{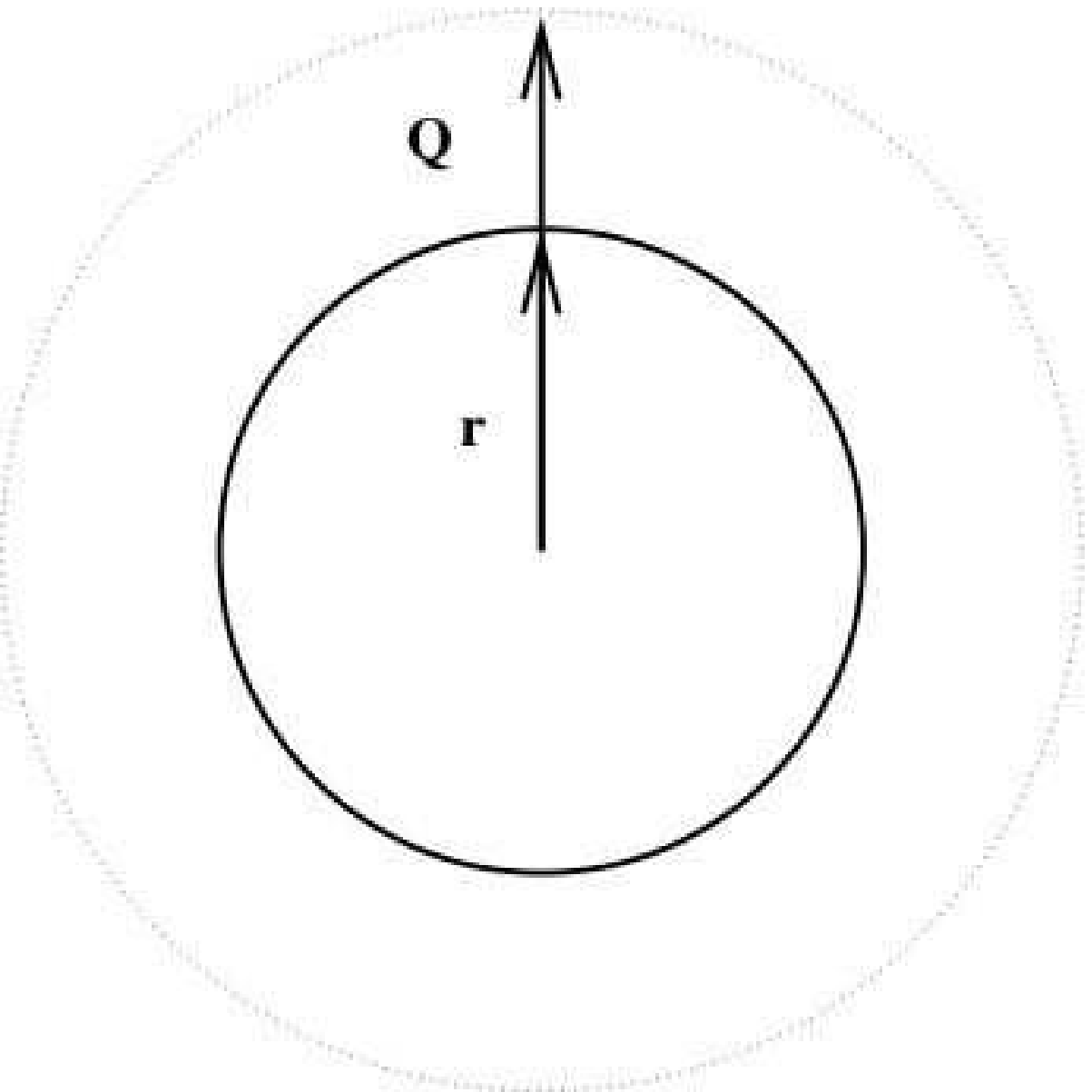}}
\vspace{-3mm} 
\parbox[t]{.55\textwidth}{\caption{In a constant electromagnetic f\/ield,
the electric and the Lorentz forces  cancel in the critical case.
The motion around the guiding center is ``frozen'' and the
particle moves according to the Hall law.\label{CFIELD2}}}
\hspace{.03\textwidth}%
\parbox[t]{.42\textwidth}
{\caption{If the electric force is harmonic, then, for critical
the magnetic f\/ield, all trajectories are circular,  and are
determined by the motion of the guiding center.\label{OSC}}}
\end{figure}

\textit{Constant fields}: ${\boldsymbol E}={\rm const}$, $B={\rm
const}$. Generically, a particle follows the usual cyclotronic
motion around the guiding center, as shown on Fig.~\ref{CFIELD2}.
${\boldsymbol q}$ is  now a constant. The velocity,
$\dot{\boldsymbol r}$, is tangent to the trajectory. It is the sum
of  the velocity of the guiding center (perpendicular to the
electric f\/ield), $\dot{\boldsymbol Q}$, and that, coming from
the rotation of ${\boldsymbol q}$~Fig.~\ref{CFIELD1}.

In the \textit{critical case} $e\theta B=1$, however (see Fig.~\ref{CFIELD2}), the electric
force is canceled by the Lorentz force:
\begin{gather*}
e\dot{{\boldsymbol r}}\times{\boldsymbol B}=e{\boldsymbol E}
\quad\Rightarrow\quad \dot{x}_i=\epsilon_{ij}\frac{E_j}{B}.
\end{gather*}

\textit{Exotic oscillator}: ${\boldsymbol E}=-\omega^2{\boldsymbol
r}$. The general motions follow elliptical orbits. In the
\textit{critical case} $\theta B=1$, however (see Fig.~\ref{OSC}), the guiding center
and the real-space position become proportional, ${\boldsymbol
q}=(1+\theta^2\omega^2){\boldsymbol r}.$ The only consistent
motions are circular, with ``Hall'' angular velocity
\begin{gather*}
\Omega=\frac{\omega^2B}{B^2+\omega^2}.
\end{gather*}

The electric force is \textit{not} compensated by Lorentz force in
this case. The dynamics is in fact \textit{non-newtonian:}
$m\ddot{{\boldsymbol r}}=\hbox{(force)}+\hbox{(terms)}$!

The reduced energy is proportional to the reduced angular
momentum,
\begin{gather*}
H_{\rm red}=\frac{\omega^2}{2}\big(1+\omega^2\theta^2\big)Q^2
\;\propto\; {\rm I}_{\rm red}=\frac{B}{2}Q^2.
\end{gather*}
The spectrum is, therefore,
\begin{gather*}
E_n=\frac{\omega^2\theta}{1+\theta^2\omega^2(1/2+n)}, \qquad
n=0,1,\dots.
\end{gather*}

\section{The semiclassical Bloch electron}\label{BlochSect}

With no relation to the above developments, a similar theory has
arisen, around the same time, in solid state physics
\cite{Niu,Niu-a}. One starts with the Bloch wave functions
\begin{gather*}
{\partial}si_{n,{\boldsymbol p}}({\boldsymbol r})=e^{i{\boldsymbol
p}\cdot{\boldsymbol r}}u_{n,{\boldsymbol p}}({\boldsymbol r}),
\end{gather*}
where $u_{n,{\boldsymbol p}}({\boldsymbol r})$ is periodic. The
vector ${{\boldsymbol p}}$ here is the crystal (quasi)momentum.
The \textit{Berry connection} is
\begin{gather*}
{\cal A}_j= i\biggl\langle u_{n,{\boldsymbol p}}\biggm|\frac{\,
{\partial} u_{n,{\boldsymbol p}}}{{\partial} p_j} \biggr\rangle.
\end{gather*}
Its curvature,
\begin{gather*}
{\boldsymbol \Theta}({\boldsymbol p})={\bf\nabla}_{{\boldsymbol
p}}\times{\cal A}_l({\boldsymbol p})
\end{gather*}
is  hence purely momentum-dependent.

Then the authors of  \cite{Niu,Niu-a} argue that  the
semiclassical equations of motion in  $n^{th}$ band should be
modif\/ied by including the Berry term, according to
\begin{gather}
\dot{\boldsymbol r}=\frac{{\partial}\epsilon_n({\boldsymbol p})}
{{\partial}{\boldsymbol p}}-\dot{{\boldsymbol
p}}\times{\boldsymbol \Theta}({{\boldsymbol p}}), \label{velrel}
\\
\dot{{{\boldsymbol p}}}=-e{\boldsymbol E}-e\dot{{\boldsymbol
r}}\times{\boldsymbol B}({\boldsymbol r}), \label{Lorentz}
\end{gather}
where ${\boldsymbol r}=(x_i)$ denotes the electron's
three-dimensional intracell position and  $\epsilon_n({\boldsymbol
p})$ is the band energy. Equations~(\ref{velrel})--(\ref{Lorentz})
derive from the Lagrangian
\begin{gather}
L^{\rm Bloch}=(p_{i}-eA_{i}({\boldsymbol r},t))\dot{x}^{i}-
(\epsilon_n({\boldsymbol p})+eV({\boldsymbol
r},t))+a^{i}({\boldsymbol p})\dot{p}_{i}, \label{blochlag}
\end{gather}
and are also consistent with the Hamiltonian structure
\begin{gather*}
\{x_i,x_j\}^{\rm
Bloch}=\frac{\epsilon^{ijk}\Theta_k}{1+e{\boldsymbol
B}\cdot{{\boldsymbol \Theta}}},
\qquad \{x_i,p_j\}^{\rm
Bloch}=\frac{\delta^{i}_{\j}+eB_i\Theta_j}{1+e{\boldsymbol
B}\cdot{{\boldsymbol \Theta}}},
\\
\{p_i,p_j\}^{\rm Bloch}=-\frac{\epsilon_{ijk}eB^k}{1+e{\boldsymbol
B}\cdot{{\boldsymbol \Theta}}}
\end{gather*}
and Hamiltonian $h=\epsilon_n+eV$
\cite{DHHMS,DHHMS-a}\footnote{This clarif\/ies the controversy
raised by Xiao et al. \cite{DHHMS-b}, see also
\cite{DHHMS-c,DHHMS-d,DHHMS-e,DHHMS-f,DHHMS-g}.}. Restricted to
the plane, these equations reduce to the exotic equations,
(\ref{DHeqmot}), when $ \Theta_i=\theta\delta_{i3}$,
$\epsilon_n({\boldsymbol p})={\boldsymbol p}^2/2m$, $
A_i=-({\theta}/{2}) \epsilon_{ij}p_j. $ Then the semiclassical
Bloch Lagrangian (\ref{blochlag}) becomes (\ref{exolag}). The
exotic Galilean symmetry is lost if $\theta$ is not constant,
though.

Recent applications of the semiclassical model include the
Anomalous \cite{AHE,AHE-a,Fang} and the spin  Hall
ef\/fects~\cite{SpinHall,SpinHall-a,SpinHall-b}. All these
developments are based on the \textit{anomalous velocity term} in
the equations of motion, $ \dot{\boldsymbol p}\times{\boldsymbol
\Theta}({\boldsymbol p}). $

\section[The anomalous Hall effect]{The anomalous Hall ef\/fect}\label{AHESect}

The anomalous Hall ef\/fect (AHE), observed in some ferromagnetic
crystals, is characterized by the absence of a magnetic f\/ield.
While it has been well established experimentally, its explanation
is still controversial. One of them, put forward by Karplus and
Luttinger \cite{Lutti}
 f\/ifty years ago, suggests that the ef\/fect is due to
an anomalous current. Here we propose to study the AHE in the
semiclassical framework.

A remarkable discovery concerns the AHE in SrRuO${}_3$. Fang et
al. \cite{Fang} have shown, by some f\/irst-principle calculation,
that the experimental data are consistent with ${\boldsymbol
\Theta}$ which behaves near ${\boldsymbol p}\approx0$ as
 \textit{monopole in momentum space}\footnote{The relevance of
the model to the AHE is still under discussion \cite{Kats}.}. Let
us consider instead the toy model given by
\begin{gather}
{{\boldsymbol \Theta}}=\theta\frac{{\boldsymbol p}}{p^3},
\label{kmonop}
\end{gather}
$p\neq0$. This  is indeed the only possibility consistent with
rotational symmetry   \cite{BeMo,PHAHE}.

For ${\boldsymbol B}=0$ and a constant electric f\/ield,
 ${\boldsymbol E}=\mbox{const}$ and assuming a parabolic prof\/ile
$\epsilon_n({\boldsymbol p})={\boldsymbol p}^2/2$,
equation~(\ref{Lorentz}) with non-commutative parameter
(\ref{kmonop}), $\dot{{\boldsymbol p}}=e{\boldsymbol E}$, is
integrated as ${\boldsymbol p}(t)=e{\boldsymbol E}\,t+{\boldsymbol
p}_0$. The velocity relation (\ref{velrel}) becomes in turn
\begin{gather}
\dot{\boldsymbol r}={\boldsymbol p}_0+e{\boldsymbol E} t
+\frac{e\theta Ek_0}{p^3}\,{\widehat{\boldsymbol n}},
\label{AHEvelrel}
\end{gather}
where  ${\widehat{\boldsymbol n}}={\widehat{\boldsymbol
p}}_0\times\widehat{{\boldsymbol E}}$ [``hats'' denote vectors
normalized to unit length]. The component of ${\boldsymbol p}_0$
parallel to ${\boldsymbol E}$ has no interest; we can assume
therefore that ${\boldsymbol p}_0$ is perpendicular to the
electric f\/ield. Writing ${\boldsymbol
r}(t)=x(t){\widehat{\boldsymbol p}}_0+y(t)\widehat{{\boldsymbol
E}}+z(t){\widehat{\boldsymbol n}}$, equation (\ref{AHEvelrel})
yields that the component parallel to ${\boldsymbol p}_0$ moves
uniformly, $x(t)=p_0t$, and its component parallel to the electric
f\/ield is uniformly accelerating,
$y(t)={\scriptstyle{\frac{1}{2}}} eEt^2$. (Our choices correspond
to choosing time so that the turning point is at $t=0$.) However,
due to the anomalous term in (\ref{velrel}), the particle is  also
deviated perpendicularly to ${\boldsymbol p}_0$ and ${\boldsymbol
E}$, namely by
\begin{gather*}
z(t)=\frac{\theta}{p_0}\frac{eEt}{\sqrt{p_0^2+e^2E^2t^2}}.
\end{gather*}
It follows that the trajectory leaves its initial plane and
suf\/fers indeed, between $t=-\infty$ to $t=\infty$,  a
\textit{finite} \textit{transverse shift}, namely
\begin{gather}
\Delta z=\frac{2\theta}{p_0}. \label{shift}
\end{gather}

\begin{figure}[t]
\centerline{\includegraphics{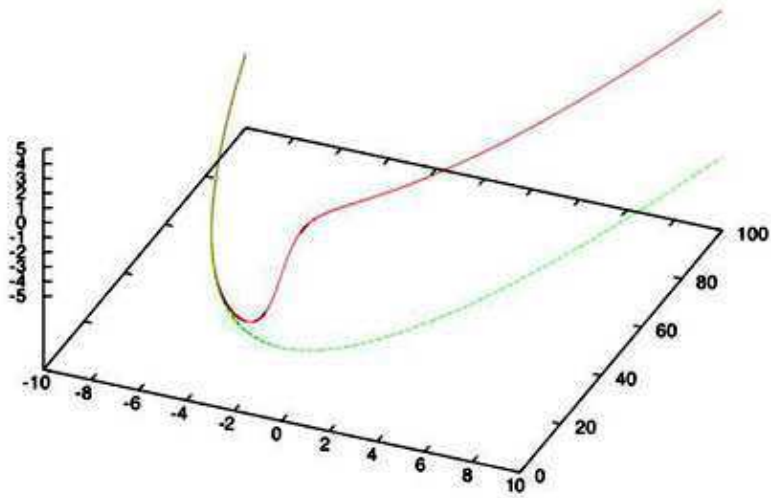}}\vspace{-3mm}

 \caption{The anomalous velocity term deviates the
trajectory from the plane. Most contribution to the shift comes
when the momentum is small, i.e., when the particle passes close
to the ``${\boldsymbol p}$-monopole''.\label{AHEplot}}
\end{figure}

Cf.~Fig.~\ref{AHEplot}, then continue with 
 $\theta$ becomes a half-integer upon quantization,
$\theta=N/2$, and hence (\ref{shift}) is indeed $N/k_0$. The
constant $p_0\neq0$, the minimal possible value of momentum, plays
the role of an impact parameter. Let us observe that while
(\ref{shift}) does not depend on the f\/ield ${\boldsymbol E}$ or
the electric charge $e$, the limit $eE\to0$ is singular. For
$eE=0$, the motion is uniform along a straight line.

The transverse shift, reminiscent of the recently discovered
optical Hall ef\/fect \cite{OptiHall},
 can also be derived  using the conservation of  angular momentum.
 The free expression\footnote{The expression (\ref{totangmom}) of the total angular momentum is \textit{not}
 mandatory, since, for a free particle, the two terms are separately conserved.}
 \cite{BeMo},
\begin{gather}
{\boldsymbol J}={\boldsymbol r}\times{\boldsymbol
p}-\theta\,{\widehat{\boldsymbol p}}, \label{totangmom}
\end{gather}
is plainly broken by the electric f\/ield
 to its component parallel to ${\boldsymbol E}$,
\begin{gather*}
J=J_y=z(t)p_0-\theta\,\frac{eEt}{\sqrt{k_0^2+e^2E^2t^2}},
\end{gather*}
whose conservation yields once again the shift (\ref{shift}).
\goodbreak

Our model is plainly not realistic: what we described is, rather,
the deviation of a freely falling non-commutative particle  from
the classical parabola found by Galileo. Particles in
a~semiconductor are not free, though, and their uniform
acceleration in the direction of ${\boldsymbol E}$ should be
damped by some mechanism.

Interestingly, a similar calculation has been performed in the
Spin-Hall context  \cite{SpinHall,SpinHall-a,SpinHall-b}. The
similarity  to optics
\cite{OptiHall,Optical,Optical-a,Optical-b,SpinOptics,SpinOptics-a}
is explained by that the three-dimensional system
 with $\Theta$ given in equation~(\ref{kmonop}) studied here is indeed
 mathematically equivalent to ``SpinOptics'', described in~\cite{SpinOptics,SpinOptics-a}.
It follows that the NC system carries therefore a massless
Poincar\'e dynamical symmetry. I am indebted to C.~Duval for
calling my attention to this point.

\LastPageEnding

\end{document}